\documentclass[twocolumn,aps,reprint,amsmath,amssymb,longbibliography,nofootinbib]{revtex4-2}
\usepackage{blindtext}
\usepackage{hyperref}
\usepackage{graphicx,color}
\usepackage[section]{placeins}
\usepackage[bottom]{footmisc}
\usepackage{float}
\newcommand{\change}[1]{{\color{black} #1}}

\usepackage{soul}

\begin{document}
\title{Multiplex mobility network and metapopulation epidemic simulations of Italy based on Open Data}
\author{Antonio Desiderio}
\affiliation{Physics Department, University of Rome “Tor Vergata”, 00133 Rome (Italy)}
\affiliation{Centro Ricerche Enrico Fermi, 00184 Rome (Italy)}
\author{Giulio Cimini}
\affiliation{Physics Department, University of Rome “Tor Vergata”, 00133 Rome (Italy)}
\affiliation{Centro Ricerche Enrico Fermi, 00184 Rome (Italy)}
\affiliation{Istituto Nazionale di Fisica Nucleare, Sezione Roma “Tor Vergata”, 00133 Rome (Italy)}
\author{Gaetano Salina}
\affiliation{Istituto Nazionale di Fisica Nucleare, Sezione Roma “Tor Vergata”, 00133 Rome (Italy)}
\date{\today}
\begin{abstract}
The patterns of human mobility play a key role in the spreading of infectious diseases and thus represent a key ingredient of epidemic modeling and forecasting. 
Unfortunately, as the Covid-19 pandemic has dramatically highlighted, for the vast majority of countries there is no availability of granular mobility data. This hinders the possibility of developing computational frameworks to monitor the evolution of the disease and to adopt timely and adequate prevention policies.
Here we show how this problem can be addressed in the case study of Italy. We build a multiplex mobility network based solely on open data, and implement a SIR metapopulation model that allows scenario analysis through data-driven stochastic simulations. The mobility flows that we estimate are in agreement with real-time proprietary data from smartphones. Our modeling approach can thus be useful in contexts where high-resolution mobility data is not available.
\end{abstract}
\maketitle
\section{Introduction}
Facing the SARS-CoV-2 outbreak has required an unprecedented effort for mankind. Governments around the world have applied containment measures in multiple occasions to hinder the spread of the virus.
In particular, as human interactions and mobility are known to deeply affect the patterns of epidemic spreading  \cite{modellin_disease_outb_urb,forecast_globalized, multiscale_mobility_vesp}, non-pharmaceutical interventions (NPIs) have been widely adopted to reduce contacts between individuals  \cite{eff_npi_haug,eff_npi_yacong,eff_npi_brito}.

NPIs can be generally divided in two main categories \cite{perra_review}. Bottom-up NPIs are self-initiated measurements and behavioral changes due to increase risk perception of the populations \cite{risk_us,risk_sk,risk_de}, which can be measured via surveys or mobile data \cite{behav_jap,behav_de,behav_smartphone}. 
Top-down NPIs are measurements imposed by governments whose effectiveness can be measured via changes in the mobility \cite{mob_rick_uk,mob_eu,npi_mob_change_ecu_col_salv} and reduction of the disease prevalence \cite{mob_and_epi_tizzoni,epi_mit_de}. 
Making informed decisions, both at the top-down and bottom-up levels, requires timely and quality data about the current stage of the outbreak and contact patterns of individuals \cite{mobile_data_informing,starnini_impact_data_accuracy}. In particular, these data serve to inform computational epidemic models \cite{castellano_vesp_review}, which produce detailed scenario analyses that can be fundamental to inform strategies of response and mitigation of a disease \cite{multiscale_mobility_vesp,real_scenario_ferguson,real_scenario_tizzoni,real_scenario_longini,real_scenario_germann} and have been widely exploited to face Covid-19 \cite{gleam_scenario,ferguson_scenario_sa,germann_scenario_us,border_scenario_us,aleta_scenario_covid,aleta_moreno_spain,mob_reop_jure_usa,scenario_barrat,scenario_susswein,scenario_aleta_moreno_vesp,scenario_colizza_did,scenario_abueg,scenario_tizz,scenario_contreras}.

Epidemic models vary in complexity \cite{castellano_vesp_review}, ranging from graph-based \cite{vesp_graph_based}, agent-based \cite{vesp_agent_based} and structured meta-population models \cite{metapop_review}, each with different data requirements. A common ingredient though is the complex network modeling of social contacts, through which contagion can spread at different scales  \cite{colizza_vesp_airline,moreno_vesp_epi_complex_networks,effect_mobility_covid_19}. 
Mobility data are crucial in this respect \cite{missing_opp}. Unfortunately, in many cases the health authorities had to deal with the scarcity and incompleteness of this type of data, which especially in the early stages of the pandemic prevented them from understanding the spreading patterns of the disease and the effects of the various containment measures implemented. 

Here we are interested in the case study of Italy, which was the first outbreak of Covid-19 among Western countries and consequently has been the subject of many research studies that tried to model and predict the evolution of the epidemics. In addition to the development of improved compartment models calibrated on the characteristics of Covid-19 \cite{epita_eq_parolini,epita_eq_giordano,epita_eq_della_rossa,epita_eq_arcede,epita_eq_de_falco,epita_a_guzzetta,epita_net_pizzuti}, much attention has been devoted to the cross-analysis of epidemic and mobility data, in order to evaluate the direct and indirect effect as well as the effectiveness of NPIs.

For instance, \cite{epita_spinella} developed a mean-field approach that allows to relate the amount of daily cases to the average number of people that an individual meets per day. 
The empirical studies have instead used mobile phone data from either Facebook Data For Good \cite{fr_it_uk_quattrociocchi,quattrociocchi_time,epita_a_bonaccorsi}
or telecom operators \cite{mob_tizzoni_italy,epita_a_pappalardo,epita_a_iacus}. 
These data are available as aggregates over territorial units, therefore they have been used to study mobility at the geographic level of regions or provinces. In particular, the authors of \cite{quattrociocchi_time} have considered the Italian regions as separate entities, due to low level of commuting between regions, while arguing that such inter-regional mobility had a crucial role in the long-range propagation of epidemics \cite{Gross_2020}. In addition, these proprietary data represent only proxies of the mobility and therefore are affected by systematic effects that cannot be controlled by researchers. 
Other works \cite{gatto_ISTAT,epita_net_parino} defined network models using commuting flows form the 2011 census database of ISTAT (the Italian Institute of Statistics), which however is obsolete and lacks information on long-range mobility.

\change{Through this work, we attempt to overcome the drawbacks of the data sources described above, by building a granular mobility network model of Italy based on data sources that are openly accessible to researchers, as well as up-to-date and complete.}
The model is defined at the detailed geographical scale of individual municipalities, and is based on a multiplex structure encompassing various means of mobility -- which can be tuned independently to mimic the implementation of different lockdown measures. 
We remark that the use of open data has benefits and drawbacks: they are readily accessible by anyone, can easily be updated in the future, 
\change{and can be tested against possible systematic errors}; however proprietary data typically have higher granularity, especially on the temporal dimension. 
\change{Besides open data, we also use a gravity-like model as described below to build the mobility network, which thus inherits all the assumption (and possible shortcomings) associated with that model \cite{barbosa_mob}.}
Nevertheless we show that the flows of population moving among provinces estimated by our multiplex model are quantitatively in agreement with those computed with Facebook Data for Good \cite{fb}. We then show how our model can be used to simulate epidemic dynamics and containment measures at the national scale. Here we consider a disease dynamics modelled via a SIR meta-population approach, but the simulation framework can be easily extended to more complex dynamics. 
We realise representative epidemic scenarios by varying the geographical position and size of the initially infected population, as well as by implementing a different combination of travel limitation measures. 
In particular we confirm that limiting long range travel is the main effective strategy to slow down the large-scale expansion of the disease. \change{Additionally we show that the effective network distance can be properly used to infer the actual seed of an epidemic outbreak.}

\section{Methods}
\subsection{Mobility network}

The mobility network is defined by representing each of the $M=7897$ municipalities in Italy as a node. We use ISTAT data (\url{https://www.istat.it/it/archivio/6789}) to collect information on identification number, reference province, population, surface, latitude and longitude of each municipality.
We consider four different mobility layers: intra-province and inter-province layers, representing short range mobility modeled with gravity-like interactions, and train and flight layers, representing long range travels extracted from the origin-destination matrices obtained by government agencies. 

To model the mobility of individuals, at each time step of the dynamics we sample a fraction $p_M = 0.1$ of the population of each municipality and move it to other municipalities. 
This moving population is distributed over the four layers with the following probabilities: $p_{\textrm{Intra}} = 0.4$, $p_{\textrm{Inter}} = 0.3$, $p_\textrm{{Train}} = 0.2$, $p_\textrm{{Flight}} = 0.1$. These values have been extracted from the origin-destination matrices of individual movements (\url{https://www.dati.lombardia.it/Mobilit-e-trasporti/Matrice-OD2020-Passeggeri/hyqr-mpe2}, as  fractions of the total moving population that uses a given transportation method or moves between different provinces. The destinations of the moving population depend on the chosen layer as described in the following. At the end of the current time step, the moving population returns to the municipality of departure.
\subsection{Short range networks: Intra-province and Inter-province layers}

The connectivity of the Intra-province and Inter-province layers is set by the territorial contiguity matrix 
collected by ISTAT (\url{https://www.istat.it/it/archivio/157423}): $c_{ij}=1$ if two municipalities $i$ and $j$ are geographically adjacent and $c_{ij}=0$ otherwise. 
We use the reference province of a municipality to split the connections into the two layers: if the province of municipality $i$ is different from the province of municipality $j$, then the connection is placed in the inter-province layer; otherwise it belongs to the intra-province layer. Connections in these layers are weighted using gravity-like assumptions \cite{barbosa_mob}: for the link between nodes $i$ and $j$ we pose
\begin{equation}
	w_{ij}^{\textrm{L}}=\frac{\rho_i\rho_j}{r_{ij}}c_{ij}^{\textrm{L}}
\end{equation}
where $\rho_i$ is the population density of node $i$ (population over surface), $r_{ij}$ is the distance between the two municipalities and $\textrm{L}\in\{\textrm{Intra, Inter}\}$ denotes the layer.
In order to transform such weights into Markovian connection probabilities we use the following normalization:
\begin{equation}
	\tilde{w}_{ij}^{\textrm{L}} = \frac{w_{ij}^{\textrm{L}}}{\sum_{l\in n_i^{\textrm{L}}} w_{il}^{\textrm{L}}}c_{ij}^{\textrm{L}}
\end{equation}
where $n_i^{\textrm{L}}$ are the territorial neighbours of node $i$ in the corresponding layer.
\subsection{Long range networks: Trains and Flights layers}

To build the Train layer, we use the RESTful API of Viaggiatreno (\url{http://www.viaggiatreno.it/}) to retrieve information of the train stations and the connections between them. Each station is identified by a unique id, latitude and longitude, while for each train departing from a given station we have the id of the next stop station and the type of connection (Frecciarossa, Intercity, Metropolitana, EuroNight, Regionale, EuroCity). To map the railway stations on the municipalities, we associated each station with the municipalities that are located no more than 5 km away (in this way, on average a single station corresponds to three municipalities). Overall the weight of a link from node $i$ to node $j$ in the train layer is built as $w_{ij}^{\textrm{Train}}=\sum_e \tau_{ij}^e g^e$, where $\tau_{ij}^e$ is a boolean indicator for the presence of a connection from $i$ to $j$ by train type $e$, while $g^e$ is the fraction of total travellers for train type $e$, obtained from ISTAT (\url{https://www.istat.it/it/archivio/13995}).
Finally, as in the short range network, to obtain connection probabilities each $w_{ij}^{\textrm{Train}}$ is normalised over the total weight of the links from $i$ to all possible destinations reachable by train in one stop.

To build the Flight layer, we collect IATA and ICAO codes, latitudes and longitudes of airports from OpenFlights (\url{https://openflights.org/data.html}. From the annual ENAC report (\url{https://www.enac.gov.it/pubblicazioni/dati-di-traffico-2019}) we collected the air traffic data: origin and destination airports, identified by the IATA and ICAO codes, and total passengers moving between airports. To map the airports on the municipalities, we associated each airport with all municipalities located no more than 15 km away (on average a single airport is mapped on ten municipalities). Overall the weight of a link from node $i$ to node $j$ in the flight layer, $w_{ij}^{\textrm{Flight}}$, is defined as the fraction of total travellers from $i$ to $j$. As for the other layers, weights are normalised to obtain connection probabilities.
\section{Results and Discussion}
\subsection{Comparison of simulated mobility fluxes with Facebook data}
\begin{figure*}[!htb]
\centering
\includegraphics[width=\textwidth]{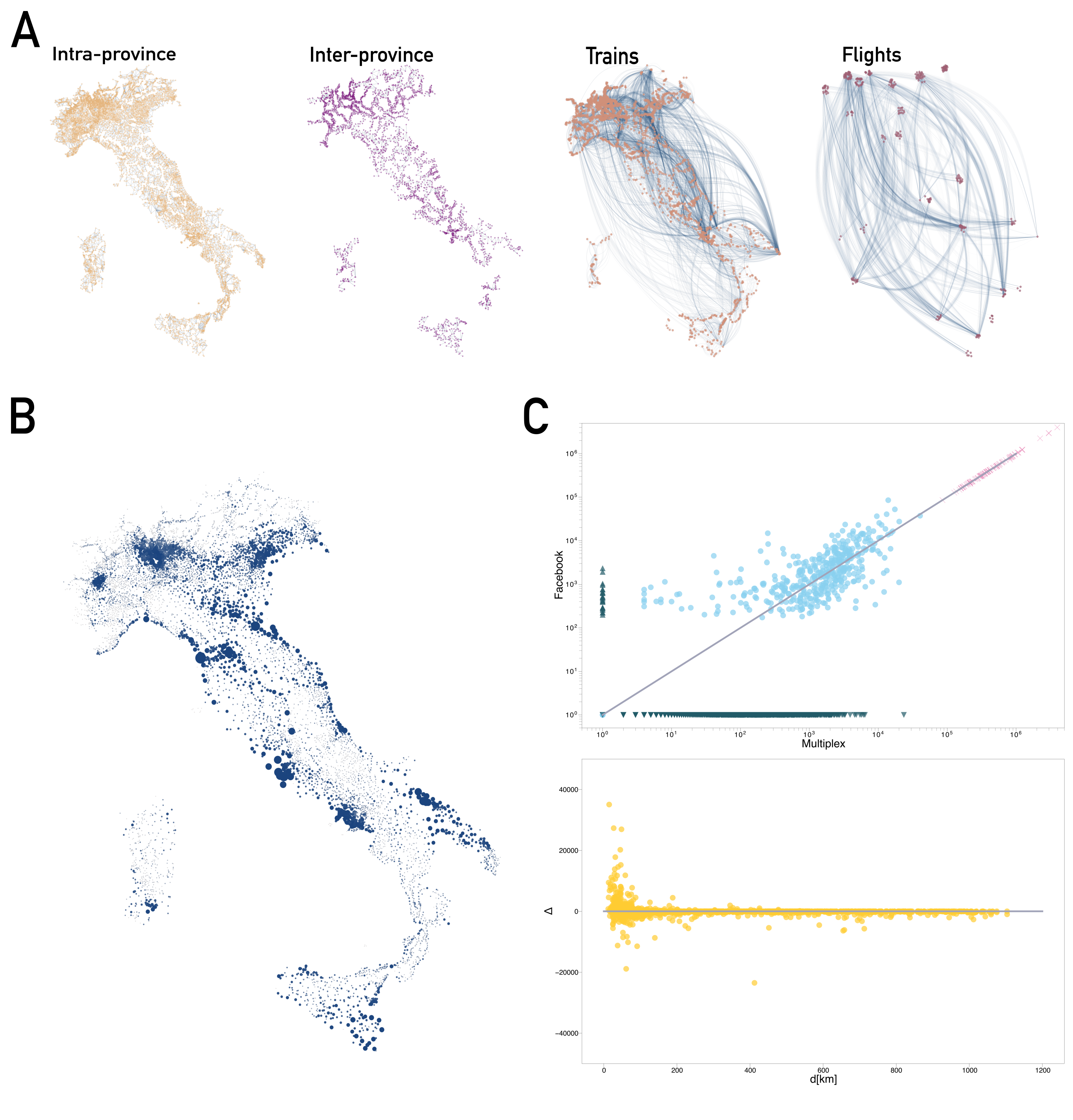}
\caption{\textbf{Multiplex mobility network of Italy.} 
A) The different layers of the mobility network of Italy, in order from \change{left to right}: intra-province, inter-province, trains and flights layer.
\change{
B) Top-ranked municipalities according to weighted k-core decomposition, with circle size proportional to the k-core value. The largest metropolitan areas (Rome, Milan, Naples, Turin, Florence, Bologna, Venice, and so on) are at the top of the ranking.
}
C) Comparison of the average population commuting among provinces between the Facebook data ( \change{baseline value}) and the model estimates for a time step of the dynamics. In the \change{upper panel} we show the scatter plot (in log-log scale) of these mobility proxies for each link, that is, each pair of provinces. Cyan dots are people moving between different provinces while pink crosses are those moving within each province. Dark green upper triangles are the links that are not present in the multiplex while the \change{(more numerous)} dark green lower triangles are those not present in the Facebook data. In the \change{lower panel} plot we show the difference of moving population between the Facebook data and the multiplex as a function of the distance between provinces.  
\change{The multiplex slightly over-estimates the long distance travels, mainly due to the many lower triangles in the upper panel, and under-estimates the short distance movements.}
}
\label{multiplex_fb}
\end{figure*}

Overall the multiplex network model is build using a combination of a gravity-like approach for short distance travels \cite{barbosa_mob} and of a data-driven approach for long range trains and flights. We have derived connection (transition) probabilities between nodes that are stationary and pairwise-independent, which allow representing mobility as a simple Markovian dynamics. 
In particular the probability that an individual moves from node $i$ to node $j$ in a single time step, $p_{ij}=p_M \sum_{\textrm{L}}p_{\textrm{L}}\tilde{w}_{ij}^{\textrm{L}}$ where $\textrm{L}\in\{\textrm{Intra, Inter, Train, Flight}\}$, is such that $\sum_j p_{ij}=1-p_{ii}$ where $p_{ii}=1-p_M$ is the probability to remain on node $i$. 
\change{Extensive probability fluxes (i.e, the moving population) between any two nodes $i,j$ can be then computed as $\omega_{ij}=P_i\,p_{ij}$, where $P_i$ is the population of the starting node $i$.} 
We remark that our model is defined at the geographical resolution of individual municipalities, contrarily to other works that take provinces or regions as minimal spatial units \cite{aleta_moreno_spain,quattrociocchi_time}. 
Figure \ref{multiplex_fb}A depicts the different layers that compose the network. 
Clearly the intra- and inter-province layers exhibit a short distance nature of the connections, whilst the trains and the flights layers represent the long range connections. 

\change{Given the mobility network, we identify the most important (or ``vital'') nodes via k-core decomposition \cite{lu_vital} based on the weighted out-degree, which for a generic node $i$ reads 
\begin{equation}
    k^W_{i} = (k_i^{a}s_i^{b})^\frac{1}{a+b}\,,
\end{equation}
with $s_i=\sum_j \omega_{ij}$ and $k_i = \sum_j \Theta(\omega_{ij})$, where $\Theta(x)$ equals 1 for $x>0$ and 0 otherwise.
We use  $a = 1/10$ and $b = 1$ to give more weight to the strength in the decomposition since it represents the extent of the mobility. Figure \ref{multiplex_fb}B shows that the top-ranked nodes are those corresponding to the largest metropolitan areas.
Notably, the weighted k-core ranking of a node is derived using only the structural properties of the network, yet it is consistent with a ranking obtained from the size of the epidemic outbreak starting from that node (Kendall's $\tau$ for the ranking restricted to region capitals equal to $\tau=0.4$ at $t=60$ time steps of the simulations and to $\tau=0.6$ at $t=80$). 
}

We then check whether our mobility model can well represent real data. We thus compare to Facebook Data for Good information; in particular we use the  \change{``baseline'' indicator of the Movement Maps. 
This value represents the daily average of the count of movements between two administrative regions over a time window of 45 days, preceding day 0 of data collection (February 23rd 2020 in our case).}
As the minimal spatial unit of the Facebook data is the province, \change{to perform the comparison} we compute inter-province fluxes in the multiplex model as the number of people commuting among provinces.  Fig \ref{multiplex_fb}\change{C} shows that such mobility fluxes are comparable to Facebook data, the main difference being the under-estimation of short distance mobility by our model and \change{the shortage} of long distance travels by Facebook data.

\subsection{Disease dynamics: different starting points}
\begin{figure*}[!htb]
\centering
\includegraphics[width=0.9\textwidth]{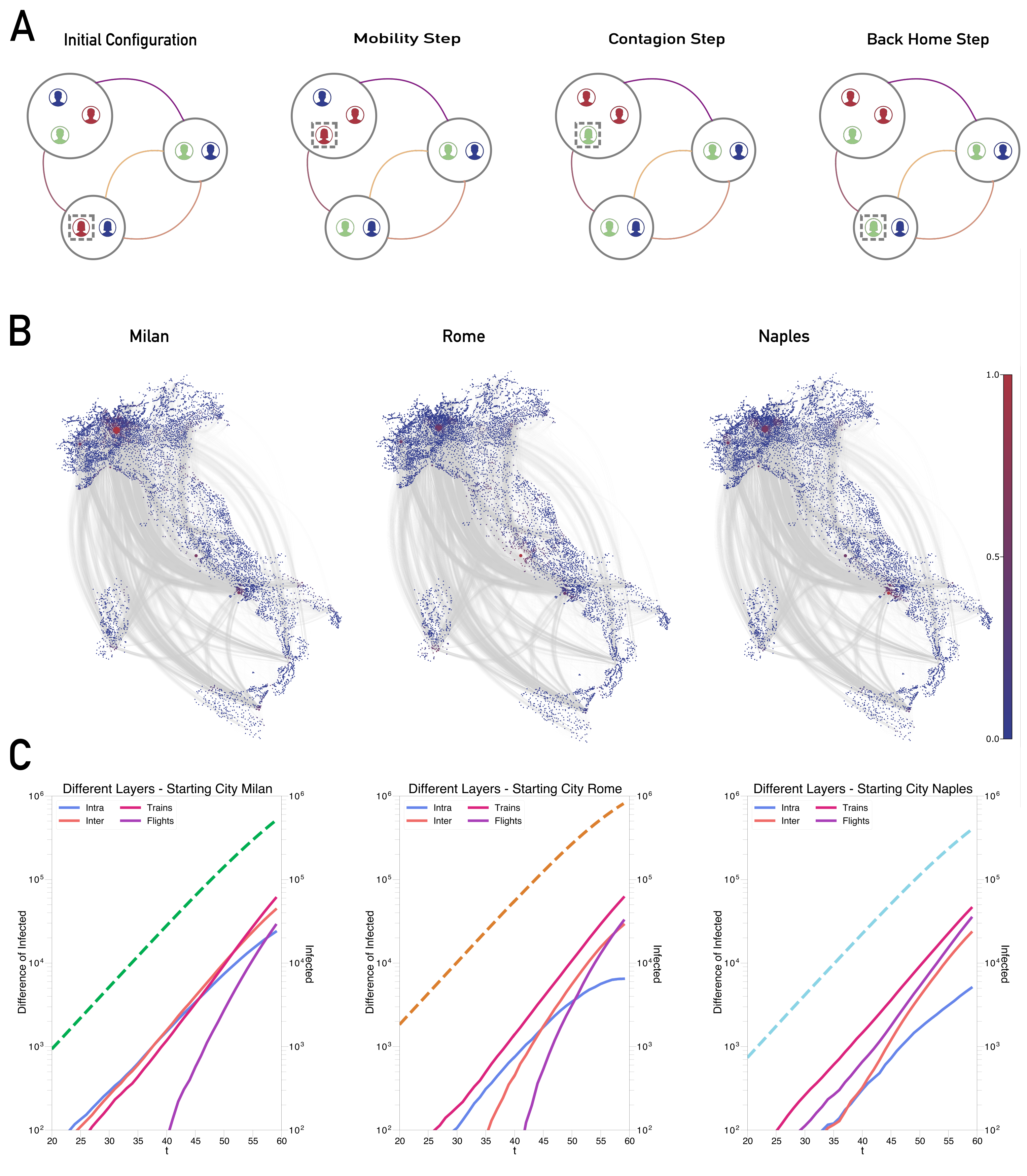}
\caption{\textbf{Epidemic \change{spreading dynamics,} scenarios for different geographic outbreaks \change{and layer importance}.}
A) Schematic representation of a single dynamical step of the model. The empty circles represents the municipalities of the network, while the edges colors are the different types (layers) of connections among them; lastly the color of the individuals represents the compartmental status: infected (red), susceptible (blue) and recovered (green). In each dynamical step we first model mobility, so that individuals move between the municipalities according to the transition probabilities, then we run disease contagion according to the SIR rules and lastly the moved individuals return to the municipality of departure.
B)\change{Snapshots} of the disease evolution over the Italian mobility network for different starting points: \change{from left to right,} Milan, Rome, Naples. In all cases the fraction of infected individuals located in the respective municipality is the same and equal to $10^{-5}$. The nodes on the maps are colored according to the normalized logarithm of the total number of infected individuals in the corresponding municipalities after 45 time steps of the dynamics.
\change{
C) Quantification of layers importance, according to the number of infected individuals for different starting points (from left to right: Milan, Rome and Naples). The dashed lines represent simulations on the full multiplex, whereas the various solid lines denote the difference with simulation outcomes performed by shutting down a given layer. 
In all cases we observe that in the early phase of the dynamics the trains layer leads to the largest difference in terms of number of infected, hence it has the largest impact. The flight layer can possibly become more important for later simulation stages.
}
}
\label{different_starting_point}
\end{figure*}

We now couple the mobility with an epidemic dynamics (see Fig \ref{different_starting_point}A) in order to simulate realistic scenarios of the disease diffusion. 
We choose to use the SIR compartmental model \change{(simulated with a discrete-time reactive process)}. 
The reason is that the simple SIR features the key dynamical ingredients that trigger the exponential trend and \change{thus it can be used to study} the territorial expansion of the disease (namely the branching process at the basis of the S$\to$I transition). Therefore this approach suffices for our purposes of simulating different scenarios on short temporal scales, without fitting of real data. 
\change{
Note however that our framework can be easily implemented with any of the more refined epidemiological models for Covid-19 proposed in the literature (see \cite{perra_review} for a review). These approach try to overcome some limitations of the SIR framework, adding for instance more compartments and dynamic values of $R_0$, as well as considering an aged-structured population.
}

Following the meta-population approach used in \cite{aleta_moreno_spain}, within a given municipality $i$ the transition rules between compartments are:
\begin{align}
    P(S_i\to I_i) &= 1 - \biggl(1 -\frac{R_0\mu}{N_i}\biggr)^{I_i}\\
    P(I_i \to R_i) &= \mu
\end{align}
where $N_i$ is the total population in municipality $i$ while $S_i$, $I_i$ and $R_i$ are the number of individuals in $i$ that are susceptible, infected and recovered, respectively. $R_0$ is the mean reproduction number and $\mu$ is the inverse of mean infectious period (here we consider values of disease parameters taken from \cite{aleta_moreno_spain}: $R_0 = 2.5$ and $\mu = 1/8$).

In Figure \ref{different_starting_point}B we plot the results of the model evolution, averaged over 100 numerical simulations, after $45$ time steps of the dynamics. We consider different \change{representative} starting \change{seeds} of the disease \change{(the seed is the municipality of the initially infected individuals)} and observe markedly different spreading patterns, both regarding the short and long distances.
\change{We also investigate the impact of each individual layer on the spreading dynamics, by removing it from the mobility network (for example, removing the Flight layer means that air travels become forbidden): individuals who are chosen to use the removed layer simply remain in their departure location. Figure \ref{different_starting_point}C shows, for the different starting seeds, the difference of infected individuals between simulations with the complete multiplex and those where a given layer is removed. We consistently observe that in the early phase of the spreading the impact of the trains layer is the largest, with the flight layer eventually becoming more important at later stages.
}

\subsection{Different scenarios of seed sizes and mobility restrictions}
\begin{figure*}[!hbp]
\centering
\includegraphics[width=\textwidth]{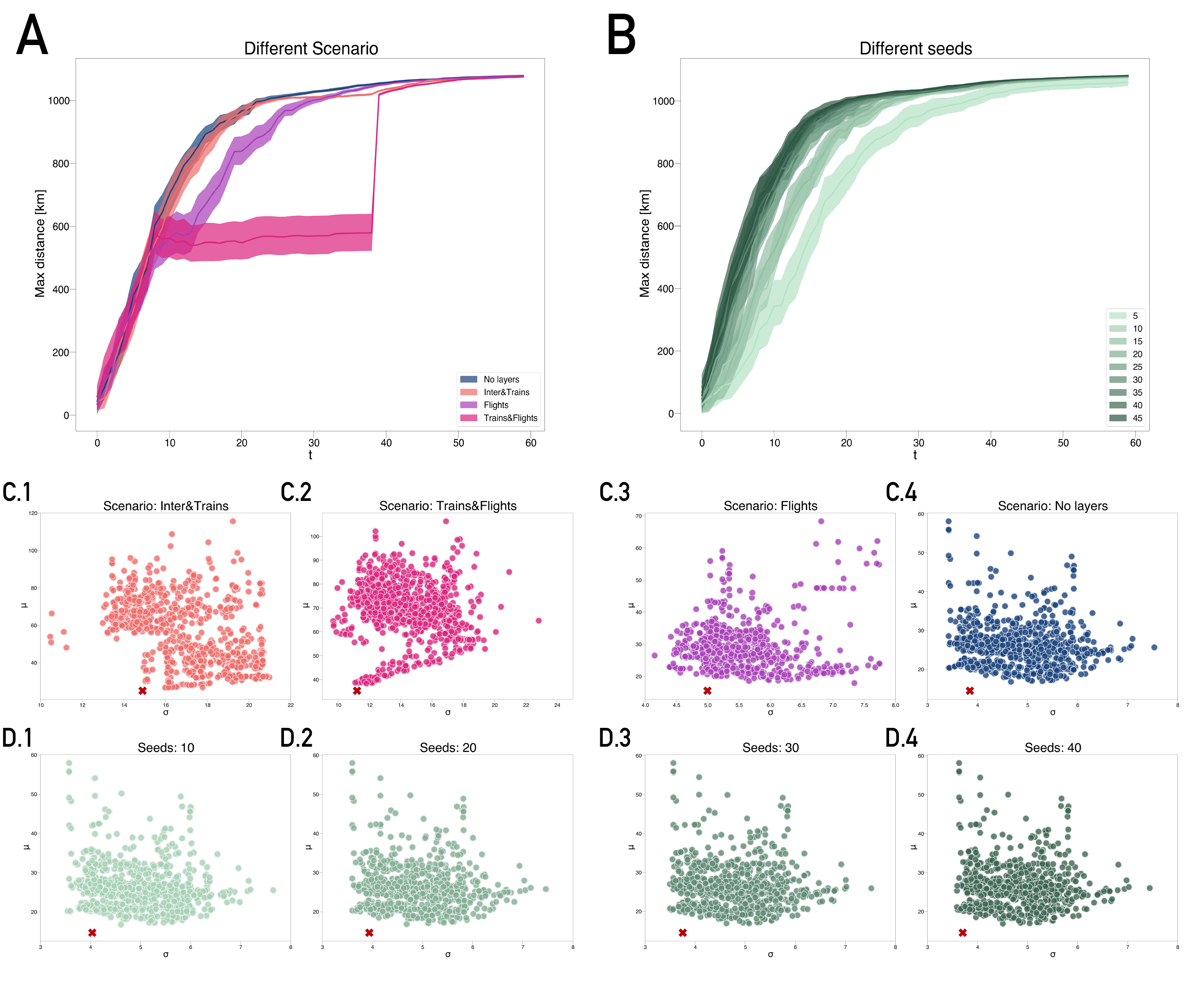}
\caption{\textbf{Epidemic scenarios for different seed sizes and travel restrictions\change{, for an outbreak starting in Milan}}.
A) \change{Maximum distance travelled by the disease} as a function of time for different mobility restriction scenarios\change{, starting with a seed size of 25 at $t=0$ (the initially infected individuals).} In a given scenario, at $t=10$ we remove all connections within a single or a pair of layers (as shown in the legend) and re-add them at $t = 40$.
B) \change{M}aximum distance reached by the disease as a function of the time steps of the dynamics, for different initial seed sizes (shown in the legend). The distance is computed using the farthest municipality with a fraction of infected individuals at least equal to that of the initial seed location at $t=0$. For each seed size we show results of 100 simulations, with solid lines representing the bootstrapped means and the shaded areas the 95\% C.I. 
\change{
C-D) Mean $\mu$ and variance $\sigma$ of effective distance of infected municipalities from potential outbreak locations at the simulation time $t=30$. The effective distance is computed considering only the mobility layers active at the given time. We consider as possible outbreak locations the top 1000 municipalities by number of populations; the red cross identifies the actual outbreak location (Milan), which is well separated from the other points.
}
}
\label{different_scenario}
\end{figure*}

Here we further illustrate how the multiplex model can be used to simulate epidemic scenarios under mobility restrictions. 
As representative example, we analyzed the maximum distance reached by the disease from the seed location. This distance is computed, for a given time step of the dynamics, considering the farthest municipality with a fraction of infected individuals at least equal to that of the initial seed location at $t=0$.
\footnote{\change{
We consider such an intensive indicator to mark a municipality as ``infected'' in order to fairly compare small and large population sizes. Other definitions are of course possible, such as those based on extensive thresholds.
}}
Figure \ref{different_scenario}B shows that this observable linearly depends on the size of the starting seed, whereas, Figure \ref{different_scenario}A illustrates the effect of different restrictions on the mobility\change{ --- }simulated by removing a combination of layers from the network. This simple exercise confirms the effectiveness of long range travels restrictions in slowing down the territorial expansion of the disease \cite{colizza_vesp_airline,Gross_2020}.

\change{
Besides geographical distance, we also consider the effective network distance as introduced in \cite{real_scenario_brockmann}: $d_{ij}=1-\log(p_{ij})$ for nodes that are directly connected, while 
$D_{ij}=\min_{\Gamma} \lambda(\Gamma)$ otherwise, where $\lambda(\Gamma)$ is the length of a directed path from $i$ to $j$, obtained by summing the effective lengths along the legs of the path. 
Figure \ref{different_scenario}C-D shows the mean and the variance of the effective distance of infected municipalities from potential outbreak locations (among the top 1000 municipalities by number of populations). 
The real outbreak location (represented as a red cross) is markedly separated from the other points. Panels C shows different mobility restriction scenarios: the mean effective distances of scenarios C.1 and C.2 (more severe restrictions) are a factor of 2 bigger than those of scenarios C.3 and C.4, and also the shapes of the point clouds are markedly different. Panels D show instead the case of different sizes of the starting seed: the variances of the effective distance, both for the actual and potential outbreak locations, decreases as the seed size increases.
}

\section{Conclusions}
In this work we developed a multiplex mobility network of Italy, defined at the fine-grained level of municipalities. We then showed how the model can be used to perform different scenario analysis. 
The Covid-19 outbreak has highlighted the lack of such a tool, at least for Italy, that would certainly have helped policy makers to monitor the disease evolution and to promptly adopt adequate prevention policies.
We built the networked data structure using only open data collected by government agencies. However the estimated mobility flows are well compatible with empirical Facebook geo-localization data from mobile phones. 
Nevertheless, the comparison with Facebook data has shown a limitation of our approach, namely the absence of a true high quality data source for short distance travels. While we used territorial contiguity to inform a gravity-like approach, considering (at the moment unavailable) open data on road and highway connections certainly represent interesting directions to further extend our approach. Overall, we believe our modeling framework can be a good starting point to build a detailed simulator for Italy to inform response strategies for future epidemics outbreaks.
\change{
\section*{Data and Code Availability}
The data are released on \href{https://zenodo.org/record/7050931\#.YxhHKC8QOL8}{Zenodo}, both in the raw scraped version and in the multiplex network construction. 
The codes to generate the mobility network and to perform the epidemic simulations are released on \href{https://github.com/RiegelGestr/multiplex_mobility_italy_municipalities}{GitHub}. For further information, contact A. Desiderio 
\href{mailto:antonio.desiderio@cref.it}{antonio.desiderio@cref.it} 
}
\change{
\section*{AUTHOR CONTRIBUTIONS}
A.D. wrote the simulation code, performed the analysis and realised the figure.
G.C. and G.S. designed the analysis and supervised the project. 
All the authors gathered the data, wrote the paper and discussed the results.
}
\change{
\section*{Competing Interest}
The authors declare no competing interests.
}

\onecolumngrid
\twocolumngrid

\onecolumngrid
\bibliography{bibliography}

\end{document}